\documentclass{PoS}

\PoS{PoS(LAT2005)144}

\usepackage{epsfig}

\title{UV-filtered overlap fermions}

\ShortTitle{UV-filtered overlap fermions}

\author{Stephan D\"urr\\
        Bern University\\
        E-mail: \email{durr(AT)itp.unibe.ch}}

\author{\speaker{Christian Hoelbling}\\
        Wuppertal University\\
        E-mail: \email{hoelbling(AT)physik.uni-wuppertal.de}}

\author{Urs Wenger\\
        NIC/DESY Zeuthen\\
        E-mail: \email{urs.wenger(AT)desy.de}}

\abstract{We discuss the kernel spectrum, locality properties and the
axial-vector renormalization constant of UV-filtered overlap fermions. We find
that UV-filtered overlap fermions have a better conditioned kernel, better
locality and an axial-vector renormalization constant closer to $1$ than their
unfiltered counterparts, even if the shift parameter $\rho$ is simply set to
$1$.}

\FullConference{XXIIIrd International Symposium on Lattice Field Theory\\
               25-30 July 2005\\
               Trinity College, Dublin, Ireland}

\newcommand{\mr}{\mathrm}
\newcommand{\gaf}{\gamma_5}

\begin{document}

%%%%%%%%%%%%%%%%%%%%%%%%%%%%%%%%%%%%%%%%%%%%%%%%%%%%%%%%%%%%%%%%%%%%%%%%%%%%%%%

\section{Introduction}

Overlap fermions \cite{Neuberger:1997fp} have many theoretically desirable
features, but they are computationally rather demanding.
There are suggestions in the literature on how to reduce the computational cost
by using a more elaborate kernel together with fattened gauge links
\cite{Bietenholz:1999km,Bietenholz:2002ks,DeGrand:2000tf}.
More recently, it has been observed that UV-filtering alone has a similar
effect \cite{Kamleh:2001ff,Kovacs:2002nz,DeGrand:2002vu,Durr:2004as,
Durr:2005an,Durr:2005ik} and maintains the $O(a^2)$ Symanzik class.
In this note we investigate the effectiveness of UV-filtering in the quenched
approximation.
We use the Wilson gauge action and employ APE \cite{Albanese:ds} or
HYP \cite{Hasenfratz:2001hp} smearing.
For details of the implementation and simulation parameters we refer the
reader to \cite{Durr:2005an,Durr:2005ik}.

%%%%%%%%%%%%%%%%%%%%%%%%%%%%%%%%%%%%%%%%%%%%%%%%%%%%%%%%%%%%%%%%%%%%%%%%%%%%%%%

\section{Kernel spectrum}

The overlap operator is constructed from the hermitean Wilson operator
$H_{\mr{W},m}=\gaf D_{\mr{W},m}$ via
\begin{equation}
D_{\mr{ov}}=\rho\left(1+\mr{sign}(H_{\mr{W},-\rho})\right)
\label{sign}
\end{equation}
where $\rho$ is the shift parameter.
In a practical implementation one has to find a method to approximate the
$\mr{sign}$ function in (\ref{sign}) over the entire eigenvalue spectrum of
$H_{\mr{W},-\rho}$.
Therefore, it is essential for an efficient implementation that the kernel
operator $H_{\mr{W},-\rho}$ is well conditioned.

\begin{figure}
\begin{center}
\epsfig{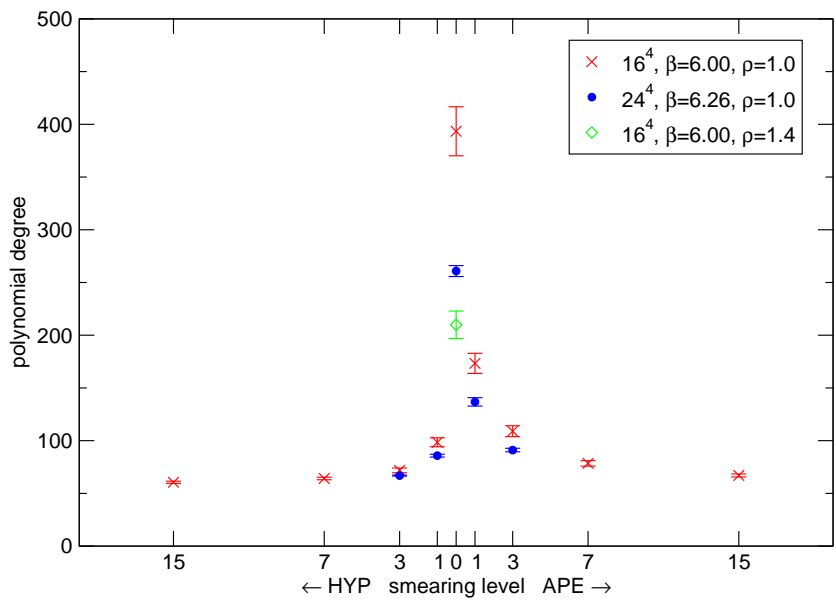}
\end{center}
\vspace{-6mm}
\caption{Degree of the Chebychev polynomial required to achieve a minimax
accuracy of $10^{-8}$ on a $16^4$ lattice at $\beta=6.0$. With $\rho=1$ a
single HYP step brings for the massless overlap operator a speedup factor
$\sim4$, but a fair comparison is to the situation with $\rho=1.4$ without
filtering, and then the factor is $\sim2$.}
\label{fig:chebychev_prac}
\end{figure}

In the free theory
%the spectrum of $H_{\mr{W},-\rho}$ is restricted to
%$[-\rho,8-\rho]$ and is empty in the interval $]-\rho,2-\rho[$.
%Therefore,
the condition number $C(\rho)$ of $|H_{\mr{W},-\rho}|$ is given by
\begin{equation}
C(\rho)=\left\{
\begin{array}{lcl}
\frac{8-\rho}{\rho}  &\;\mr{for}& 0<\rho\le1\\
\frac{8-\rho}{2-\rho}&\;\mr{for}& 1\le\rho<2
\end{array}
\right.
\label{condnrfree}
\end{equation}
which has a minimum value 7 at $\rho=1$.
In the interacting theory the upper edge of the spectrum at $8-\rho$ is
only slightly affected for typical $\beta$.
On the other hand the absence of a lower bound on the spectrum can result in a
dramatic increase of the condition number $C(\rho)$.
Since there are standard techniques to treat the lowest few of these eigenmodes
exactly, the relevant issue is the onset of the bulk of these modes, and it
turns out that the latter gets significantly lifted by UV-filtering
\cite{Durr:2005an}.

In order to translate the condition number into computational cost, we choose
to present the degree of the Chebychev polynomial \cite{Hernandez:2000sb}
necessary to reach a given precision in the minimax norm after treating the
$14$ lowest eigenmodes exactly.
As one can see from Fig.\,\ref{fig:chebychev_prac}, the number of forward
applications of $H_{\mr{W}}$ needed to obtain a given precision is
significantly reduced by UV-filtering.

\begin{figure}
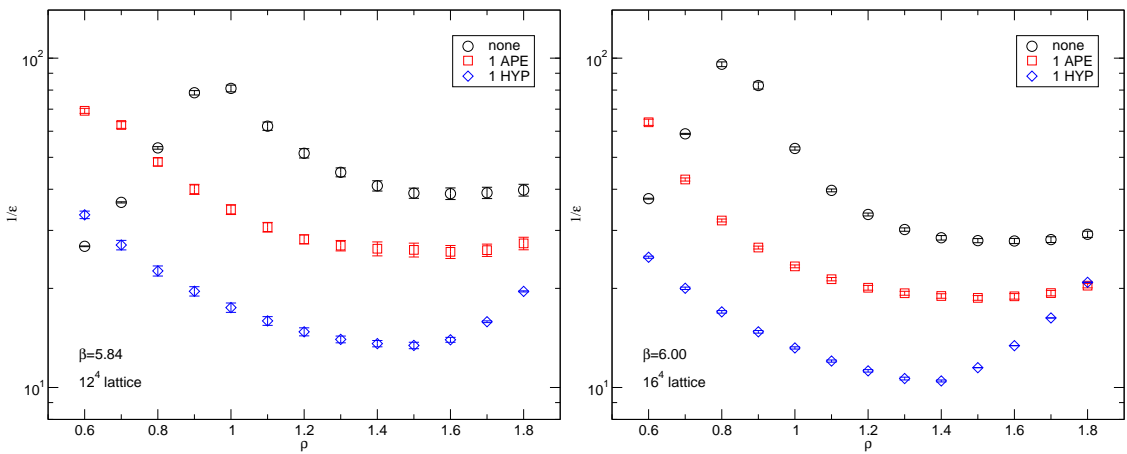

\begin{center}
\epsfig{file=condition_number_L12_b5.84.eps,width=7.4cm}
\epsfig{file=condition_number_L16_b6.00.eps,width=7.4cm}
\end{center}
\vspace{-6mm}
\caption{Condition number $1/\epsilon=C(\rho)$ of $|H_{\mr{W},-\rho}|$ versus
$\rho$ for $\beta=5.84$ (left) and $\beta=6.0$ (right).}
\label{fig:cond}
\end{figure}

More details are shown in Fig.\,\ref{fig:cond}.
A rather conservative filtering recipe like applying 1 step of APE link
fattening seems to render the kernel far more benevolent; the condition number
remains comfortably small for a large range of shift parameters, including the
canonical choice $\rho=1$.

%%%%%%%%%%%%%%%%%%%%%%%%%%%%%%%%%%%%%%%%%%%%%%%%%%%%%%%%%%%%%%%%%%%%%%%%%%%%%%%

\section{Locality}

The ``spread'' inherent in the gauge link fattening raises the question whether
a UV-filtered overlap operator is less local than the original thin link
version.
It has already been observed in \cite{Kovacs:2002nz} that the UV-filtering
actually proves beneficial to the locality properties of the overlap operator.

\begin{figure}
\begin{center}
\epsfig{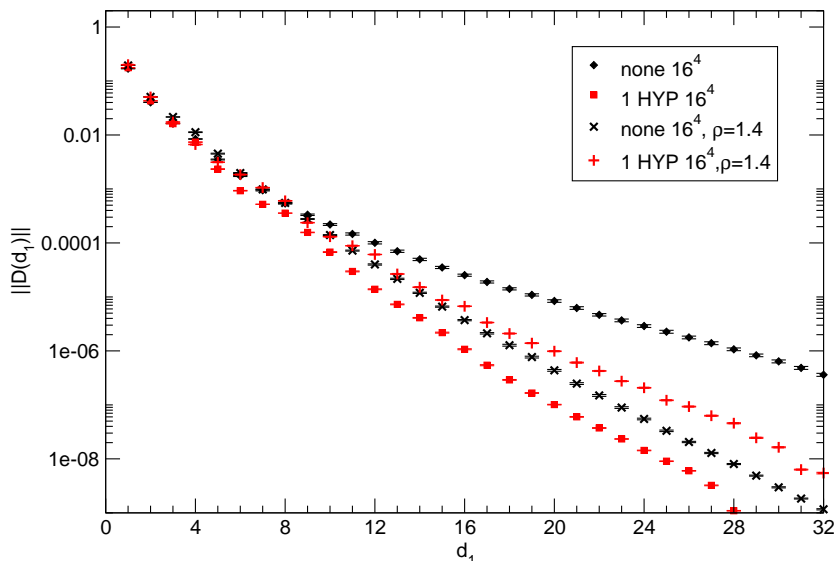}
\end{center}
\vspace{-8mm}
\caption{Localization of the overlap operator at $\beta=6.0$ without filtering
and after 1\,HYP step, for $\rho=1$ and $\rho=1.4$. A single HYP step
proves more efficient than optimizing $\rho$. Filtering and $\rho>1$ should
not be combined; for the filtered operator the untuned choice $\rho=1$ is
reasonable (but still not optimal).}
\label{fig:localization_6.0}
\end{figure}

Fig.\,\ref{fig:localization_6.0} shows the localization of the overlap operator
at $\beta=6.0$.
Following the convention of \cite{Hernandez:1998et} we plot the maximum over
the 2-norm of $D_\mr{ov}\eta$ at $x$ with $\eta$ a normalized $\delta$-peak
source vector at the point $0$ in the lattice versus the ``taxi driver''
distance $d_1=||x||_1$ to the location of the $\delta$-peak
\begin{equation}
f(d_1)=\mr{sup}\left(||(D_\mr{ov}\eta)(x)||_2\Big| ||x||_1=d_1\right)
\;.
\label{supremum}
\end{equation}
Looking first at the unfiltered operators (black/dark diamonds and crosses) one
finds the well known result that (at this $\beta$) adjusting $\rho$ to a value
around $1.4$ lets $f(d_1)$ fall off steeper than keeping the untuned value
$\rho=1$.
The interesting observation is that a single HYP step together with $\rho=1$
(red/light squares) results in an even steeper descent than the unfiltered
version with $\rho=1.4$ (which was chosen to nearly optimize the locality of
the unfiltered operator).
The last curve shown (red/light pluses) indicates that one should not attempt
to combine the filtering with a $\rho$ value that would be optimal for the
unfiltered operator.

%%%%%%%%%%%%%%%%%%%%%%%%%%%%%%%%%%%%%%%%%%%%%%%%%%%%%%%%%%%%%%%%%%%%%%%%%%%%%%%

\section{Renormalization}

One interesting aspect of UV-filtered fermions is their improved
renormalization behavior.
It has been realized some time ago \cite{Bernard:1999kc,DeGrand:2002va} that in
general one-loop corrections to renormalization constants are substantially
smaller in the case of UV-filtered fermions.

\begin{table}[!b]
\begin{center}
\begin{tabular}{|l|cccc|}
\hline
~  &  $5.66$ &  $5.76$ &  $5.84$ &  $6.00$ \\
\hline
$Z_A^\mr{none},\rho=1$   &ill-def.&ill-def.&7.06(73)&3.145(94)\\
$Z_A^\mr{none},\rho=1.4$ \cite{Wennekers:2005wa}&---&---&$\sim1.71$&1.553(02)\\
$Z_A^\mr{none},\rho=1.6$ \cite{Babich:2005ay}   &---&---&$\sim1.44$& ---\\
$Z_A^\mr{1\,APE}\,,\rho=1$&2.57(7) & 1.90(2) & 1.66(02) & 1.452(04)\\
$Z_A^\mr{3\,APE}\,,\rho=1$&1.55(3) & 1.33(1) & 1.23(01) & 1.160(06)\\
$Z_A^\mr{1\,HYP},\rho=1$  &1.44(2) & 1.28(1) & 1.22(01) & 1.153(03)\\
$Z_A^\mr{3\,HYP},\rho=1$  &1.21(1) & 1.13(1) & 1.10(01) & 1.072(02)\\
\hline
\end{tabular}
\end{center}
\vspace{-4mm}
\caption{$Z_A$ with several filterings. Note that
\cite{Wennekers:2005wa,Babich:2005ay} give $Z_A$ values at slightly different
$\beta$ only. The precise numbers are
$Z_A^\mr{none}(\beta=5.8458,\rho=1.4)=1.710(5)$ \cite{Wennekers:2005wa} and
$Z_A^\mr{none}(\beta=5.85,\rho=1.6)=1.443(5)$ \cite{Babich:2005ay}.}
\label{tab:Z_A}
\end{table}

To investigate this behavior in a nonperturbative context we choose to
determine the axial-vector renormalization constant $Z_A$ from the AWI,
following the method of \cite{Luscher:1996sc,Giusti:2001pk}.
Our results are collected in Tab.\,\ref{tab:Z_A} \cite{Durr:2005an}.
It is interesting to note that for the untuned $\rho=1$ thin link operator
$Z_A$ is large even at $\beta=6$.
With $\rho=1.4$ and the same coupling $Z_A$ is much closer to $1$, indicating
that $\rho>1$ is not only essential for the locality of the unfiltered
$D_\mr{ov}$ but also beneficial to its renormalization constants.
Note, however, that even the 1-fold APE smeared operator with the untuned
$\rho=1$ has an axial-vector renormalization constant which is closer to $1$.
When going to the coarser $\beta=5.84$ lattice $Z_A$ of the thin link operator
with $\rho=1$ increases drastically.
At this point it becomes essential for the thin link operator to properly tune
$\rho$.
Going to even coarser lattices we were unable to obtain a signal for $Z_A$ with
the unfiltered operator and $\rho=1$ (for details see \cite{Durr:2005an}).

\begin{figure}
\begin{center}
\epsfig{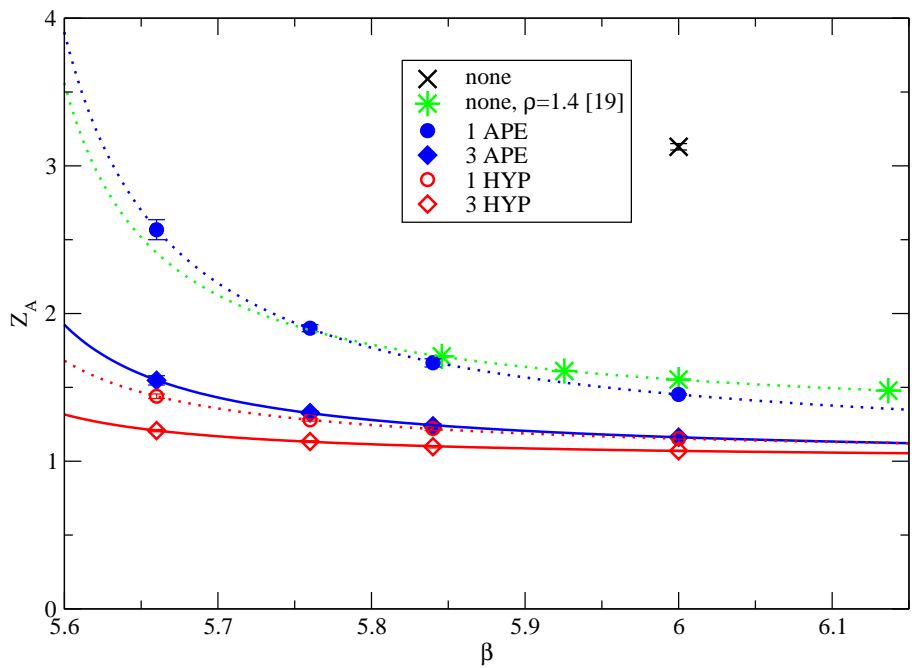}
\end{center}
\vspace{-6mm}
\caption{Axial-vector renormalization constant $Z_A$ versus $\beta$. The thin
link data at $\rho=1.4$ are from \cite{Wennekers:2005wa}.}
\label{fig:zap}
\end{figure}

\begin{table}
\begin{center}
\begin{tabular}{|l|ccccc|}
\hline
filtering level  & none, $\rho=1.4$ \cite{Wennekers:2005wa}
& 1 APE &  3 APE & 1 HYP & 3 HYP \\
\hline
$c_1+c_3$ & 1.66(16) & 0.36(32) & 0.03(20) & 0.15(15) & 0.03(13) \\
$c_3$ & 5.54(4) & 5.53(2) & 5.51(4) & 5.49(4) & 5.49(5) \\
\hline
\end{tabular}
\end{center}
\vspace{-4mm}
\caption{The 1-loop coefficient $c_1+c_3$ and pole location $c_3$ of a fit
(4.1) for different levels of UV-filtering. Note, that the analytic value for
$c_1+c_3$ for thin link overlap at $\rho=1.4$ is $c_1+c_3=0.722408$
\cite{Alexandrou:2000kj,Capitani:2000da}.}
\label{tab:padec}
\end{table}

In Fig.\,\ref{fig:zap} the $Z_A$ with different filterings are plotted vs.\
$\beta$.
The lines represent fits of the form
\begin{equation}
Z_A(x)=\frac{1+c_1 x+c_2 x^2}{1-c_3 x}
\label{zap}
\end{equation}
with $x=1/\beta$ and all fits have a reasonable $\chi^2$.
To compare to the thin link $\rho=1.4$ case we use data from 
\cite{Wennekers:2005wa}.
In principle, these curves contain two pieces of information.
The asymptotic slope $c_1+c_3$ for $x\to 0$ predicts the perturbative 1-loop
coefficient for $Z_A$ and the pole in (\ref{zap}) [i.e.\ the value of $c_3$]
predicts the coupling where the ansatz (\ref{zap}) breaks down.
With the current quality of our data all asymptotic slopes --~except for the
thin link operator~-- are compatible with zero and we are unable to make a
quantitative statement about the perturbative 1-loop coefficients
(see Tab.\,\ref{tab:padec}).
Still, on a qualitative level the filtered one-loop coefficients are smaller
than both the analytic result and the fit to nonperturbative data for thin
link overlap with $\rho=1.4$.
This suggest a much better behaved perturbative series for the filtered
overlap operator \cite{Durr:2005an}.
For the location of the pole the results in Tab.\,\ref{tab:padec} indicate
that this quantity is hardly changed by the filtering.
This suggests that the (practical) range of validity of the perturbative
expansion is barely affected by the filtering \cite{Durr:2005an}.

%%%%%%%%%%%%%%%%%%%%%%%%%%%%%%%%%%%%%%%%%%%%%%%%%%%%%%%%%%%%%%%%%%%%%%%%%%%%%%%

\section{Conclusions}

We have presented some evidence that UV-filtered overlap fermions might bring
substantial technical advantages over their thin link counterparts.
In particular, a forward application of the massless overlap operator requires
fewer applications of the hermitean Wilson operator, the locality of the
overlap operator is generally improved with UV-filtering and the axial-vector
renormalization constant is much closer to $1$.
In addition, there is no need to tune the kernel shift parameter $\rho$.
The canonical choice $\rho=1$ is satisfactory even on rather coarse lattices.

%%%%%%%%%%%%%%%%%%%%%%%%%%%%%%%%%%%%%%%%%%%%%%%%%%%%%%%%%%%%%%%%%%%%%%%%%%%%%%%

\end{document}